\documentclass[12pt]{article}
\begin{document}
\baselineskip 18pt
\newcommand{\Tr}{\mbox{Tr\,}}
\newcommand{\Dirac}{/\!\!\!\!D}
\newcommand{\beq}{\begin{equation}}
\newcommand{\eeq}[1]{\label{#1}\end{equation}}
\newcommand{\bea}{\begin{eqnarray}}
\newcommand{\eea}[1]{\label{#1}\end{eqnarray}}
\renewcommand{\Re}{\mbox{Re}\,}
\renewcommand{\Im}{\mbox{Im}\,}
%\begin{titlepage}
\begin{titlepage}
\hfill  CERN-TH/2002-176 hep-th/0207255
\begin{center}
\hfill
\vskip .4in
{\large\bf OBSERVATIONS ON THE HOLOGRAPHIC DUALS OF 4-D EXTREMAL BLACK HOLES}
\end{center}
\vskip .4in
\begin{center}
{\large S. Ferrara$^{a,b}$ and M. Porrati$^{c,d}$}\footnotemark
\footnotetext{e-mail: sergio.ferrara@cern.ch, massimo.porrati@nyu.edu}
\vskip .1in
(a){\em Theory Division CERN, Ch 1211 Geneva 23, Switzerland}
\vskip .1in
(b){\em INFN, Laboratori Nazionali di Frascati, Italy}
\vskip .1in
(c){\em Department of Physics, NYU, 4 Washington Pl.,
New York, NY 10003, USA}
\vskip .1in
(d){\em Rockefeller University, New York, NY
10021-6399, USA}
\vskip .1in
\end{center}
\vskip .4in
\begin{center} {\bf ABSTRACT} \end{center}
\begin{quotation}
\noindent
We argue that extremal black holes of N=4 Poincar\'e supergravity coupled 
to conformal matter have a quantum-exact dual 5-d description in the 
maximally supersymmetric extension of the Randall-Sundrum theory. 
This dual is a the classical, static supergravity solution describing a 
string with both Neveu-Schwarz and Ramond 
charge with respect to different antisymmetric tensor doublets. We also 
discuss the issue of the singularity present in a class of such supergravity
solutions found by Cveti\v{c}, L\"u and Pope.
\end{quotation}
\vfill
\end{titlepage}
\eject
\noindent
\section{Introduction}
The 5-d Randall-Sundrum (RS) compactifications~\cite{rs1,rs2}  
have a holographic 
interpretation in terms of a dual 4-d theory consisting of dynamical gravity
coupled to a strongly-interacting field theory (see \cite{ahpr} and references 
therein; see also~\cite{rz,dl}). Its supersymmetric extension has been
studied by many authors. In particular, the findings in~\cite{dls,lp,clp}
will have a direct bearing on this paper. 
The N=4 extension of the model in~\cite{rs2}, usually called RS2, 
allows for compelling 
quantitative checks due to non-renormalization theorems~\cite{gk,dl}. 
An interesting problem in this context is to find the 5-d configuration dual to
a 4-d Schwarzschild 
black hole on the UV brane. The linearized analysis of~\cite{gkr} has not
been extended to an exact all-order solution, in spite of various attempts in 
the literature. 

Recently, an intriguing reason for this failure has been 
proposed by Emparan, Fabbri and Kaloper~\cite{efk}, who have pointed out that
the 5-d configuration dual to the Schwarzschild black hole must also 
reproduce its Hawking radiation {\em at the classical level}, so that it 
cannot be stationary, since the 4-d black hole necessarily radiates away 
in an asymptotically Minkowsky background.

In any extension of the RS2 scenario with N$\geq 2$ supersymmetry, the 4-d
dual possesses 
absolutely stable extremal black holes, that preserve some of the
supersymmetry and do not Hawking-radiate. This suggests the possibility that
such 4-d black holes can be represented in 5-d $AdS$ space 
by classical, {\em static} 
configurations. Indeed, solutions were found in~\cite{lp} in the N=2
case, and in~\cite{clp} in the N=4 case. They are strings ending on the UV 
brane, and extending in a straight line from the brane to the 
$AdS_5$ horizon~\footnote{By $AdS_5$ horizon we mean, of course, the horizon 
of the Poincar\'e coordinate patch.}. 
Their drawback is that they always have null singularities on the $AdS_5$
horizon.
In this paper, we use general properties of maximally extended supersymmetry
to argue that, whether or not the solutions in~\cite{lp,clp} are unique, 
{\em any} other solution corresponding to an extremal 4-d black hole of
N=4 Poincar\'e supergravity must describe a string ending on the UV brane.
and it must also be quantum-mechanically  stable. 

In the next Section, we briefly recall some relevant aspects of 4-d,
N=4 Poincar\'e supergravity, and its extremal black holes. In Section 3
we identify the duals of these black holes in N=8, $AdS_5$ 
supergravity~\cite{grw,ppvn}. 
In particular, we point out that the duals are strings carrying
charges under some of the 12 antisymmetric forms of the 5-d gauged 
supergravity, which are doublets of $SL(2,R)$ and in the 6 of 
$SU(4)\sim SO(6)$. 
The fact that the antisymmetric forms are doublets of $SL(2,R)$ means
that they define charges of both Ramond and Neveu-Schwarz type. 
If the string carries a single charge, it is 1/2 supersymmetric; otherwise, it
is 1/4 supersymmetric. This property has a nice holographic counterpart. The 
antisymmetric forms are dual to the 6 graviphotons of the 4-d Poincar\'e
theory, and 4-d black holes can carry both electric and magnetic charges 
under these $U(1)$s. Black holes carrying only one type of charges are 1/2 BPS,
while those with both electric and magnetic charge (under different $U(1)$s)
are 1/4 BPS~\cite{kp,klopvp}. 
This fact is yet another proof that the $SL(2)$ electric-magnetic 
duality in 4-d, N=4 supergravity~\cite{csf} maps into the $SL(2)$ symmetry of
type IIB superstrings. 
We conclude in Section 4 with some remarks on the puzzle arising from
 the explicit form of the known 5-d supegravity metrics associated with 
BPS black holes.
\section{N=4, 4-d Poincar\'e Supergravity and Its Extremal Black Holes}
N=4 Poincar\'e Supergravity in four dimensions~\cite{bks} 
has two types of multiplets: gravitational, and matter. 

The gravitational multiplet contains 16 bosons plus 16 fermions:
\beq
g_{\mu\nu},\qquad A_\mu^{[AB]}, \qquad S, \qquad \psi^A_{\mu\alpha},
\qquad \chi_A.
\eeq{1}
Here $A,B=1,..4$ are $SU(4)$ indices. Notice that $[AB]\sim \Lambda$, where
$\Lambda=1,..6$ is a vector index of $SO(6)\sim SU(4)$. 
$S=i\exp(-2\phi) +a$, $\phi$ is the string dilaton, and $a$ is the RR 10-d 
axion.

Matter, which in what follows play no significant role, is in
N=4, Lie-algebra valued Yang-Mills multiplets. In the solutions we will
discuss, these fields are set consistently to zero. 

The N=4 pure supergravity black holes are the axion-dilaton black holes 
studied in~\cite{kp,klopvp,ko,o}.  In particular, 1/4 BPS black holes have
nonzero entropy, 
given by the Bekenstein-Hawking formula, and their axion-dilaton 
scalar evolves towards a fixed value at the horizon~\cite{fks,fk1,fk2}.
In a generic BPS black hole background of pure N=4 supergravity, the central
charge (complex) $SO(6)$ vector is 
\beq
Z_\Lambda=e^{K/2} (q_\Lambda - S p_\Lambda),
\eeq{2}   
where $q_\Lambda, p_\Lambda$ are the asymptotic charges coupled to the 6
graviphotons. $K=-\log (i\bar{S} - iS)$ is the K\"ahler 
potential of the $SU(1,1)/U(1)$ axion-dilaton sigma model.
$Z_\Lambda$ changes by an overall phase under the $SL(2,Z)$
symmetry of type IIB 10-d supergravity.

The eigenvalues $Z_1,Z_2$ of the central-charge matrix are given 
by~\cite{fk2,dlr}
\bea
|Z_{1,2}|^2 &=& {1\over 2} (|Z_1|^2 + |Z_2|^2) \pm {1\over 4} \sqrt{
p^2 q^2 - (p\cdot q)^2}, \nonumber \\
|Z_1|^2 + |Z_2|^2 &=& {1\over 2} Z_\Lambda \bar{Z}^\Lambda =
{1\over 4}[e^{2\phi} q^2 + e^{-2\phi} p^2 + e^{2\phi} (a^2 p^2 -2a p\cdot q )]
\eea{3}
The modulus $|Z_1|$ is the ADM mass (in Planck units) of the extremal 
black hole. If $p^2q^2= (p\cdot q)^2$, $|Z_1|=|Z_2|$, and the resulting 
configuration is 1/2 BPS and has zero entropy. If $p^2q^2> (p\cdot q)^2$,
the resulting configuration is 1/4 BPS, and it has a nonzero entropy given by
the attractor equation~\cite{fks,s1,fk1,fk2}
\beq
S= {1\over 4}A = \pi | Z_1^*|^2,
\eeq{4} 
where $ Z_1^*$ is the value of $Z_1$ at the stationary point
\beq
{\partial \over \partial a} Z_1 =  {\partial \over \partial \phi} Z_1 =0.
\eeq{5}
These equations fix the value of the dilaton  and the axion to
\beq
a^*={p\cdot q \over p^2}, \qquad e^{-2\phi^*}= {1\over \sqrt{p^2}} \sqrt{ q^2
-{(q\cdot p)^2\over p^2}}.
\eeq{6}
Substituting these values into Eq.~(\ref{4}), we arrive at
\beq
S= {1\over 2}\pi \sqrt{p^2 q^2 -(p\cdot q)^2}.
\eeq{7}

We note that by an $SO(6)$ rotation, followed by an $SL(2,R)$ transformation,
we can set $q_\Lambda=(q,0,0,...0)$, $p_\Lambda=(0,p,0...,0)$, which gives
$\exp(-2\phi^*)=|q/p|$, $a^*=0$. Then, we can re-write the equation for 
$|Z_1|$ as
\beq
|Z_1|= {1\over 2\sqrt{2}} (e^{\phi} |q| + e^{-\phi} |p|).
\eeq{8}
\section{The 5-d Duals of 4-d Extremal Black Holes}
The 1/2 and 1/4 BPS black holes we have briefly described here are
extremal, so that they do not decay by Hawking radiation. Since they also
possess some residual supersymmetry, they belong to short multiplets of
N=4, and are thus completely stable, both classically and quantum mechanically.
Therefore, their 5-d duals must be classical, 
static configurations of $AdS_5$ supergravity.

To identify these configuration, we must first understand which 5-d fields 
give rise to the graviphotons. 
The crucial observation is that they {\em cannot} 
be the $SU(4)$ 5-d gauge fields, since they
have no zero mode, and, moreover, they do not transform under the $SU(4)$ 
symmetry as the 4-d graviphotons. As we noticed previously, the graviphotons 
of 4-d Poincar\'e supergravity belong to the 6 of $SO(6)$.
Moreover, their corresponding electric and magnetic fields are doublets under
$SL(2,R)$. These are precisely the quantum numbers of the 12 antisymmetric
tensors of the gauged N=8 supergravity in $AdS_5$~\footnote{The antisymmetric
tensors obey first-order equations, so that they can be naturally identified 
with 4-d field strengths, as it is required by the $SL(2,R)$ 
symmetry~\cite{dls}.}. It is therefore natural 
to identify the 4-d graviphotons with the zero modes of these antisymmetric 
tensors. This identification is indeed the correct one, as shown 
in~\cite{lp,dls} for the N=4, $AdS_5$ reduction, and in~\cite{clp} for the N=8 
reduction.

Since antisymmetric fields couple to strings, instead of point particles,
it is obvious that the 5-d configuration dual to extremal black holes 
{\em must} be strings. 
The ones corresponding to regular, nonzero-entropy black holes must
carry charge under at least two antisymmetric forms, one of NS and the other of
R type. 

The form of the N=8 Poincar\'e superalgebra, and its possible $AdS_5$
extension, also implies that the 5-d strings are in the 6 of $SO(6)$~\cite{fp}.
To see this, we write the relevant part of the 4-d super-Poincar\'e algebra,
namely the anticommutator of two same-chirality supercharges 
\beq
\{Q^A_\alpha,Q^B_\beta\}=\epsilon_{\alpha\beta}Z^{[AB]},\qquad 
\alpha, \beta=1,2.
\eeq{8a}
This expression corresponds to the 5-d anticommutator 
\beq
 \{Q^A_a,Q^B_b\}=(\gamma_\mu C)_{ab} Z^{\mu\,[AB]}, \qquad a,b=1,..4, \qquad
\mu=0,..4.
\eeq{8b}
This equation shows as promised that 5-d strings are in the antisymmetric
representation of $SU(4)$.

Supergravity solutions corresponding to 5-d string have been found 
in~\cite{clp}. In that paper, it
was shown that it is possible to lift the $SL(2,Z)$ black holes of N=4
Poincar\'e supergravity to full solution of the equations of motion of
N=8, $AdS_5$ supergravity 
(indeed, to solutions of 10-d type IIB supergravity). 
The significance of those solutions has been questioned, since they exhibit
null singularities at the $AdS_5$ horizon.
 
What is clear from our analysis of the R-symmetry of the 4-d Poincar\'e
theory, is that any 5-d configuration corresponding to a BPS black hole must be
a string, perpendicular to the UV brane. The fact that these configurations
have a source extending into the 5-d bulk may seem strange; nevertheless, it
does not contradict any property of the holographic duality, and
the R-symmetry properties of the graviphotons require it to be so. 
Before discussing further the explicit solutions of refs.~\cite{lp,clp}, we
must notice that Eq.~(\ref{5}) can be exactly 
reinterpreted as the extremization of the 5-d string tension~\cite{adfl}, 
since, 
schematically
\beq
L\times\mbox{String Tension in 5d} \sim \mbox{BH Mass in 4-d}.
\eeq{9}
Here $L$ is the radius of curvature of $AdS_5$.  

In order to have 1/4 BPS states it is essential that the 5-d two-forms 
are nonsinglets under the R-symmetry. Therefore, unlike the case of 10 
dimensions~\cite{s}, $(p,q)$-strings can also be 1/4 BPS, instead of 
just 1/2 BPS.

We conclude this Section by pointing out that unlike 5-d Poincar\'e 
supergravity, $AdS_5$ supergravity does not allow for duality transformations
exchanging antisymmetric forms with vectors.
\section{The Singularity at the IR Horizon}
The known black strings solutions corresponding to 4-d, 1/4 BPS black holes 
have 5-d metric and dilaton given by~\cite{clp}
\bea
ds^2 &=& {L^2\over z^2}\left[ dz^2 -(H_1H_2)^{-1} dt^2 + H_1 H_2(dr^2 +
r^2 d\Omega_2^2)\right], \nonumber \\
e^{-2\phi} &=& {H_1\over H_2}, \qquad H_1= {1\over g_S} +  {p\over r} ,
\qquad H_2=g_S + {q\over r}.
\eea{10}
Here $d\Omega^2_2$ is the standard metric element on the  unit-radius 
2-sphere, and $g_S$ is the string coupling constant, i.e. the asymptotic value
of $\exp(\phi)$ away from the string. 

This metric has a null singularity at $z=\infty$~\cite{clp}. Since we  
argued in the previous Sections that {\em all} 1/4 BPS black holes have
stable, classical $AdS_5$ duals describing strings extending along the
coordinate $z$ through the horizon at $z=\infty$, we are faced
with two possibilities: either there exist more complicated string 
solutions besides Eq.~(\ref{10}), or the null singularity at $z=\infty$ 
can be made meaningful in the context of the holographic duality.

Since Eq.~(\ref{10}) admits Killing spinors~\cite{lp,clp}, it seems at least
plausible
that it is unique. As we have argued, since the 5-d solution describes a 
string, it must have anyway a cylindrical horizon, extending all the way to 
$z=\infty$.
The possibility exists, therefore, that whatever the 5-d duals of 1/4 BPS 
black holes are, they do have a naked singularity at $z=\infty$. 
That this is not a disaster may be argued using the holographic duality.
In the 4-d picture, indeed, the null singularity describes a pathology 
arising only in the extreme infrared region of the theory. Because of the 
UV/IR duality, $z=\infty$ corresponds to infinite wavelength. This means that
the singularity may simply reflect a pathology unaccessible to any local 
observer. We may also recall that even 
worse naked singularities do arise in N=8,
$AdS_5$ supergravity solutions describing RG flows to non-conformal field
theories (see for instance~\cite{gppz,g}). These singularities reflect the 
existence of physics beyond the supergravity approximation, rather than a
pathology of the RG flow itself. Indeed, in some cases, their resolution is
explicitly known~\cite{ks,jpp}. 
\vskip .1in 
\noindent
{\bf Acknowledgements}\vskip .1in
\noindent
We would like to thank M.J. Duff, 
R. Emparan, J. Fernandez-Barbon, R. Rattazzi and 
R. Sundrum for useful discussions. M.P. would like to thank CERN for its kind 
hospitality. M.P. is supported in part by NSF grant no. PHY-0070787.
S.F. is supported in part by the European Community's Human Potential Program 
under contract HPRN-CT-2000-00131 Quantum Space-Time, and by DOE under grant
DE-FG03-91ER40662, Task C.

\end{document}